# Modeling time-resolved kinetics in solids induced by extreme electronic excitation

N. Medvedev[1,2,*], F. Akhmetov[3], R.A. Rymzhanov[4,5], R. Voronkov[6], A.E. Volkov[6]

1) Institute of Physics, Czech Academy of Sciences, Na Slovance 2, 182 21 Prague 8, Czech Republic
2) Institute of Plasma Physics, Czech Academy of Sciences, Za Slovankou 3, 182 00 Prague 8, Czech Republic
3) Industrial Focus Group XUV Optics, MESA+ Institute for Nanotechnology, University of Twente, Drienerlolaan 5, 7522 NB Enschede, The Netherlands
4) Joint Institute for Nuclear Research, Joliot-Curie 6, 141980, Dubna, Moscow Region, Russia
5) The Institute of Nuclear Physics, Ibragimov St. 1, 050032 Almaty, Kazakhstan
6) P.N. Lebedev Physical Institute of the Russian Academy of Sciences, Leninskij pr., 53,119991 Moscow, Russia

## Abstract

We present a concurrent Monte Carlo (MC) – molecular dynamics (MD) approach to modeling of matter response to excitation of its electronic system. The two methods are combined on-the-fly at each time step in one code, TREKIS-4. The MC model describes arrival of irradiation, which in the current implementation can consist of a photon, an electron, or a fast ion. It also traces induced cascades of excitation of secondary particles, electrons and holes, and their energy exchange with atoms due to scattering. The excited atomic system is simulated with an MD model. We propose a simple and efficient way to account for nonthermal effects in the electron-atom energy transfer in covalent materials *via* conversion of potential energy of the ensemble into the kinetic energy of atoms, which can be straightforwardly implemented into an MD simulation. Such a combined MC-MD approach enables us time-resolved tracing of the excitation kinetics of both, electronic and atomic systems, and their simultaneous response to a deposited dose. As a proof-of-principle example, we show that proposed method describes atomic dynamics after X-ray irradiation in a good agreement with tight-binding MD, with much more affordable computational demands. The new model also allows us to gain insights into behavior of the atomic system during the energy deposition from a nonequilibrium electronic system excited by an ion impact.

## 1. Introduction

Various types of irradiation causing initial extreme electronic excitation are employed for research and practical applications: photon, electron, and ion beams. Creating a highly excited state of matter [1,2], femtosecond X-ray free-electron laser (FEL) pulse irradiation is used to process materials for nanotechnology, in biophysics researches, and for production and studies of exotic states of matter such as warm dense matter [1–5]. Providing molecular imaging with unprecedented time-resolution on the order of a few femtoseconds, two colors FEL beams are applied for investigations of temporal kinetics of the fundamental processes in matter and molecules under extreme electronic excitations [3–5].

---

[*] Corresponding author: ORCID: 0000-0003-0491-1090; Email: nikita.medvedev@fzu.cz

Irradiation with electrons is used in studies of material properties. It has a practical interest for electron microscopy, where electrons irradiating the target being imaged lead to degradation of the material [6]. Ultrafast electron diffraction recently found many practical applications, allowing to achieve femtosecond resolution to study fundamental effects in matter [7,8].

Swift heavy ions (SHI) impacts modify materials within nanometric proximity of their trajectory, which makes them unique tools for nanotechnological applications [9,10]. Irradiation with heavy ions is applied for creation of nano-pores and membranes, used e.g. for medical diagnostics [11]. SHIs are used in biology and medicine for tumor treatments [12,13] and modeling of cosmic rays effects [14].

Transient processes in highly excited targets push standard models beyond their limits and motivate development of new approaches [15], forming fundamental interest for basic research. All these applications, resulting from the rapid and extreme excitation of the electronic system, require detailed modeling for deeper understanding and control of the occurring processes.

Excitation of an electronic system by any of the abovementioned sources triggers a sequence of processes, starting with elastic and inelastic scattering of fast electrons. The elastic scattering provides energy exchange with target atoms without secondary electron excitation. The inelastic one generates new electrons and holes in the core shells or the valence/conduction band of the material [15,16]. Typically, such processes occur at femtosecond timescales, cooling electrons within sub-picoseconds. Upon relaxation of the electronic system, the target atoms react to the energy transferred at pico- to nano-second timescales [17,18]. Finally, after cooling of the atomic system, the target may be left in a structurally modified state, observable experimentally [1,9].

In the last decade, it became apparent that the standard calculations of the total deposited dose are not sufficiently precise to describe the local material modifications. This methodology of the total dose calculations stems from radiobiology, satisfactorily describing macroscopic effects. Nanotechnological applications (such as quantum dots [19,20], nano-electronics [21], nanomembranes [22,23]), or precise control of biological damage (such as DNA modifications [24,25]) require more detailed understanding of the microscopic processes involved. The quest for ever-increasing precision inevitably leads to the quest for deeper understanding of the fundamental processes of damage formation caused by the electronic excitation, and development of appropriate models [18].

This stimulated development of hybrid and multiscale models that are becoming more and more common in applications. Such models combine a few different approaches into one code tracing various aspects of the problem [18]. Each individual model covers its own characteristic scales, for example, ultrashort time-scales may be described with precise *ab-initio* approaches, whereas macroscopic timescales may be traced with more approximate continuous (thermodynamic) models [18]. In a more general case, in a hybrid approach, each model describes its own processes under appropriate conditions. For example, different subsystems (electrons and ions) may be described with different methods.

In the next section, we will describe TREKIS-4, the hybrid approach developed here, placing it into the context of modern existing hybrid and multiscale codes. We present a novel approach to coupling of the electrons and atoms in simulations. The simple, efficient and precise scheme proposed enables advances in modeling of materials response to various types of radiation causing a strong primary disturbance of the electronic system.

We demonstrate examples of application of this approach to description of an early stage of a material response to an X-ray pulse and swift heavy ion irradiation. The combined approach allows us to gain an insight into the stage of overlapping electronic and atomic kinetics, which was not possible to obtain with previous state-of-the-art approaches, proving the versatility and capability of the proposed methodology.

## 2. Model

TREKIS-4 (which stands for Time-Resolved Kinetics in Irradiated Solids) is a hybrid model combining two different methodologies on-the-fly: Monte Carlo (MC) model with molecular dynamics (MD) simulations. The Monte Carlo module traces: (i) the incident particle penetration (which may be a photon, an electron, a positron, or an ion, in the current implementation); (ii) transport of all secondary excited electrons; (iii) kinetics of electronic holes; and (iv) transport of emitted photons. The MD module of TREKIS-4 simulates the dynamics of target atoms triggered by the energy deposition from the electrons and holes. Such a combination allows tracing all essential processes, responsible for damage formation in a target under irradiation, e.g. phase transitions under femtosecond X-ray pulse irradiation, or formation of nanometric tracks after penetration of a swift heavy ion, as will be discussed in the Results section.

Earlier attempts combined MC and MD models sequentially: output data from MC simulations of electronic kinetics were passed as initial conditions to MD modeling of atomic response [26–28]. In contrast, TREKIS-4 executes both models in parallel. As will be described below, the data are exchanged on each time step of the simulation. The idea is similar to two temperature model (TTM) combinations with MD, where the state of the electronic system is simulated with temperature diffusion equation, while the atomic system is traced with MD [29,30]. TTM-MD is currently a standard tool for simulation of laser-irradiated damage in metals, but it proved insufficient for modeling of SHI impacts and resulted in the need to use adjustable parameters: most often, the electron-phonon coupling coefficient is fitted to reproduce experimental data on the track radius [31]. It is also more complicated to apply TTM-MD to nonmetallic targets [30,32,33]. In contrast to TTM, Monte Carlo modeling is fully capable of tracing non-equilibrium and ballistic electronic transport, effects of valence holes in non-metallic materials, core holes kinetics, transport of photons, and it does not require fitting parameters [34].

A similar idea for a concurrent MC-MD scheme was previously used in a few codes, such as e.g. MBN Explorer [35], XTANT [36,37], XMDYN [38]. MBN Explorer uses MD simulation of atomic motion in biological samples, and a kinetic MC simulation of various chemical processes. It does not explicitly trace secondary electron and holes transport, approximating electronic processes with rate equations instead. XMDYN code combines molecular dynamics simulations of both, ions and electrons in plasma created by irradiation of samples with femtosecond X-ray pulses, with Monte Carlo simulation tracing atomic processes such as Auger and radiative cascades. It assumes a plasma limit, tracing only highly energetic particles, and not processes in solid targets such as phase transitions. XTANT code combines tight-binding MD with MC tracing of X–ray photons and nonequilibrium electrons, and Boltzmann scattering integrals for slow, equilibrium fraction of electrons.

The major differences between TREKIS-4 and XTANT are in the two main aspects: a) TREKIS-4 relies on the classical MD model, whereas XTANT uses tight-binding MD; b) TREKIS-4 has much more advanced MC simulation module, accounting for many processes that are not included in XTANT. Point (a) allows TREKIS-4 to simulate much larger systems, with spatial dimensions that XTANT cannot cover. Point (b) makes TREKIS-4 more universal, allowing for modeling of a wide variety of incident radiation types and conditions (ions, electrons, positrons, photons, including relativistic energies) vs. only X-ray photons in non-relativistic energy range in XTANT. On the other hand, since XTANT does not rely on the classical MD potentials, it is capable of tracing evolution of the electronic structure and interatomic forces due to electronic excitation (such as nonthermal melting [37]). Thus, the codes serve different purposes and are complementary to each other.

Below we describe all the details and parameters of the models implemented in TREKIS-4: the Monte Carlo method for each type of simulated particles, the molecular dynamics approach for target atoms, and the interplay between the two modules. The Monte Carlo simulation carries all the iterations in parallel, discretized in the same time steps as MD to be able to gather and average information from all iterations at each time step (see Chapter 15 in [18]).

The transport MC module in TREKIS-4 uses analog (event-by-event) algorithm for propagation of classical trajectories of individual particles [18]. A particle free flight between sequential scattering events, $\lambda$, is approximated as a straight line and is sampled according to the exponential distribution:

$$\lambda = -\lambda_0 \ln(\xi_1), \qquad (1)$$

here $\xi_1 \in [0,1)$ is a uniformly distributed random number, and the mean free path $\lambda_0$ is defined *via* a cross section, $\sigma$, as follows [39,40]:

$$\lambda_0^{-1} = \sigma n_{at} = n_{at} \int_{W_-}^{W_+} \int_{Q_-}^{Q_+} \frac{d^2\sigma}{dWdQ} dWdQ, \qquad (2)$$

here $n_{at}$ is the density of target atoms; $\frac{d^2\sigma}{dWdQ}$ is the double differential cross section, $W_\pm$ and $Q_\pm$ are the minimal and maximal transferred energy ($W$) and momentum in the energy units ($Q(1 + Q/2m_t c^2) = \hbar^2 q^2/(2m_t)$, where $q$ is the transferred momentum and $m_t$ is the scattering center mass, $\hbar$ is the Planck constant, and $c$ is the speed of light in vacuum), [41]. All masses used are rest masses of particles.

For an incident particle with the kinetic energy $E$ and mass $M$, the integration limits are defined as follows [41]:

$$\begin{cases} W_- = I_p \\ W_+ = \frac{2m_t c^2 E(E + 2Mc^2)}{2m_t c^2 E + (Mc^2 + m_t c^2)^2} \end{cases}, \qquad (3)$$

$$Q_\pm = \sqrt{\left(\sqrt{E(E + 2Mc^2)} \pm \sqrt{(E-W)(E-W+2Mc^2)}\right)^2 + (m_t c^2)^2} - m_t c^2,$$

where $I_p$ is the ionization potential of the atomic shell an electron is being ionized from in a case of inelastic scattering, $m_t$ is the mass of a target scattering particle. In a case of ionization from a valence band, $I_p = E_{gap}$ ($E_{gap}$ is the band gap of the material). In a case of scattering on a conduction band electron, or elastic scattering on atoms, $I_p = 0$. In a case of identical particles scattering (electron-electron), $W_+ = (E + I_p)/2$ [41].

In a scattering event, the sampled transferred energy, $\delta E$, is also defined by the differential scattering cross section according to the following expression [36]:

$$\xi_2 \sigma = \int_{W_-}^{\delta E} \int_{Q_-}^{Q_+} \frac{d^2\sigma}{dWdQ} dWdQ \qquad (4)$$

here $\xi_2 \in [0,1)$ is a different random number. Eq.(4) is solved for the transferred energy $\delta E$ with the bisection method (if an analytical solution is not available). Then, a particle deflection angle is calculated from the momentum conservation prior to subtracting ionization potential (in the case of inelastic scattering) [42].

The following sections describe the specifics for each type of particle modeled and further simulation details.

## 2.1. Cross sections in the linear response approximation

In the linear response approximation, the cross section of a charged particle scattering reduces to the dependence on the longitudinal energy loss function (ELF) – an imaginary part of the inverse complex dielectric function (CDF) of a target, $Im(-1/\varepsilon(W, Q))$ [41,43]:

$$\left(\frac{d^2\sigma}{dWdQ}\right)_l = \frac{Z^2 Z_t^2}{n_{at}\pi a_0 m_e c^2 \beta^2}\left[1 - \exp\left(-\frac{W}{T}\right)\right]^{-1} \times \frac{2Mc^2}{WQ(Q+2Mc^2)} W \frac{Q + m_t c^2}{m_e c^2} Im\left(\frac{-1}{\varepsilon(W,Q)}\right), \qquad (5)$$

Here $Z$ and $Z_t$ are the charges of the incident particle and the target particle (scattering center), which will be discussed below for each type of particles; $a_0$ is the Bohr radius; $\beta = v/c = \sqrt{1-(1+E/Mc^2)^{-2}}$ is the incident particle velocity normalized to the speed of light in vacuum; $m_e$ is the free-electron mass; and $T$ is the target temperature (in energy units).

In the ultra-relativistic case, a transverse part of CDF starts to contribute [16]. This contribution is currently neglected in TREKIS-4, which sets its upper limit of applicability at mildly relativistic energies [41].

With all the other parameters known, the problem reduces to evaluation of the CDF. Here, we use the Ritchie-Howie formalism to obtain the CDF function parameters, approximated with the Drude-Lorentz oscillators form [44]:

$$Im\left(\frac{-1}{\varepsilon(W,Q)}\right)_{RH} = \sum_{i=1}^{N_{osc}} \frac{A_i \gamma_i W}{\left[W^2 - (E_{0_i}+Q)^2\right]^2 + (\gamma_i W)^2} \quad (6)$$

where $\gamma_i$ are the widths of the oscillators, $E_{0_i}$ define their positions, and $A_i$ are their weights or contributions of target electrons into this oscillator. The set of parameters $\{\gamma, E_0, A\}_i$ for the number of oscillators $N_{osc}$ is fitted to reproduce available experimental or calculated optical parameters in the optical limit ($Q=0$) [42,45].

It is known that in the case of protons transport, Ritchie-Howie oscillators in CDF do not produce sufficiently accurate energy losses [46,47]. In this case, Mermin-like CDF oscillators can be used, following Ref. [46,47]. It also may improve other particles mean free paths calculations at low energies, if required. Mermin-like CDF uses the same input from the optical coefficients (optical limit Eq.(6)), but the extension from the optical limit to finite $Q>0$ is done via Mermin dielectric function.

At high energies of a projectile, when ($\{\gamma_i\}\ll\{W,Q\}$), the CDF may be reduced to a sum of the Dirac's delta-functions by formally setting $\gamma_i \to 0$ [41]:

$$Im\left(\frac{-1}{\varepsilon(W,Q)}\right) = \sum_{i=1}^{N_{osc}} \frac{\alpha_i(0)}{W}\delta(W-(E_{0_i}+Q)) \quad (7)$$

here $\alpha_i(0) = \pi A_i/2$ obtained from the same optical coefficients as Eq.(6). Integration with the Dirac's delta-function alters the integration limits (3) to more complicated expressions defined in [41], which are, however, straightforward to implement into a numerical program.

Ritchie-Howie CDF coefficients for a number of solids can be found, e.g., in Ref. [42]. For other materials, one may construct CDF oscillators for a wide list of materials using optical coefficients from [48,49] (for low-energy photons) or databases [50,51] and from [52,53] (for high-energy photons), following the detailed instructions in Ref. [54].

In an absence of optical parameters, the single-pole approximation may be used. In this case, Eq.(6) is approximated with a single oscillator $N_{osc} = 1$ [55]. The parameters of the CDF may then be approximated as follows. The position of the oscillator may be chosen according to the position of a collective mode of the particles oscillations: phonon mode for elastic scattering on the atomic system of a target (here approximated with the Einstein oscillation frequency, $\omega_{ph}$), or the plasmon mode for inelastic scattering on the valence/conduction band electronic system:

$$\begin{cases} E_{0_{sp}} = \hbar\omega_{ph} = \pi\hbar c_s \sqrt[3]{n_{at}}, & \text{elastic} \\ E_{0_{sp}} = \max(\hbar\Omega_p, I_p), & \text{inelastic} \end{cases} \quad (8)$$

Here, $c_s$ is the speed of sound in the material, and the plasmon frequency is defined as $\Omega_P^2 = 4\pi e^2 n_e/m_e$, with $n_e$ being the valence/conduction band electron density. The choice of the maximum between the plasmon energy and ionization potential automatically identifies the plasmon energy for the valence/conduction band scattering, and ionization potential for core-shell scattering.

Within the single-pole approximation the width of the CDF peak is currently empirically approximated as $\gamma_{sp} = E_{0_{sp}}/10$ for elastic scattering and $\gamma_{sp} = E_{0_{sp}}$ for inelastic one (which enter only Eq.(6) and vanish for delta-functional approximation Eq.(7)). The normalization coefficient is unambiguously defined by the *k*-sum rule [56]:

$$A_{sp} = (\hbar\Omega_p)^2 / \int_0^\infty W Im\left(\frac{-1}{\varepsilon(W, Q=0, A=1)}\right)_D dW \quad (9)$$

This rather crude approximation allows evaluating the scattering cross section and mean free paths of charged particles in complex materials, for which the optical coefficients are unknown.

To summarize, in TREKIS-4, the following cross-sections are used for scattering of charged particles: wherever available, nonrelativistic inelastic scattering is described with Ritchie-Howie CDF Eq.(6), except for protons, for which Mermin-like CDF is used. For high-energy particles, delta-functional approximation Eq.(7) is used. For materials, for which the optical coefficients are unavailable, the single-pole approximation with the parameters (8)-(9) is employed.

## 2.2. Ions

TREKIS-4 is capable of modeling ion impacts, from protons up to super-heavy ions, in the regime of the electronic energy loss [57]. In this regime, the elastic (nuclear) scattering of a projectile is negligible, and only inelastic scattering is considered, since it forms the main contribution into the ion energy loss ($S_e >> S_n$, where $S_e$ is the electronic energy loss, and $S_n$ is the nuclear energy loss) [42]. Other processes such as radiation emission *via* Bremsstrahlung and Cherenkov emission are also negligible for ions since Bremsstrahlung is inversely proportional to the mass of a projectile [58]. So, only the inelastic scattering of a projectile is considered, resulting in impact ionization of electrons in target atoms or valence/conduction band.

To describe the inelastic scattering, cross sections from Eqs. (1-5) are used, with the CDF models mentioned in the previous section. In the case of scattering on target electrons, the target mass and charge are set as free electron parameters: $m_t=m_e$, $Z_t=1$. The ion charge for protons is set $Z=1$, whereas for heavier particles (referred to as swift heavy ions, SHI), we use the Barkas approximation for the effective charge, $Z=Z_{eff}(v)$ [59]:

$$Z_{eff}(v) = Z_{ion}\left[1 - \exp\left(-\frac{v}{v_0}Z_{ion}^{-\frac{2}{3}}\right)\right], \quad (10)$$

where $v$ is the SHI velocity, $v_0=c/125$ is empirically adjusted atomic electron velocity [59], and $Z_{ion}$ is the atomic number of the SHI in the Periodic Table. Such a simple approximation allows us to describe SHI energy loss across the entire energy range from a few MeV to TeV, without tracing complex charge exchange processes between the SHI and a target [41,42,60].

### 2.3. Electrons

Whether we are interested in an external electron impact on a target, or transport of secondary electrons excited by a projectile, modeling of electrons in a wide range of energies from a few eV up to MeVs must include at least the following processes: elastic and inelastic scattering, Bremsstrahlung photon emission, and boundary crossing with electron emission from the surface (if free surfaces are included in the simulation) [16]. For elastic and inelastic scattering, we use the linear response theory (Eqs.(1-7)), with the incident electron mass and charge set to free particle parameters $M=m_e$, $Z=1$.

For inelastic scattering, the target mass and charge are also set as free electron parameters: $m_t=m_e$, $Z_t=1$. In an inelastic scattering event, a new electron is created, leaving a hole in a deep atomic shell, or in the valence/conduction band of the material. The shell or band being ionized is chosen according to the partial ionization cross section, normalized to the total one [40,42]. In the case of the valence band ionization, the energy state, from which an electron is being excited, is chosen according to the unperturbed density of states (DOS) in the initial and the final states of the material, for a given sampled transferred energy $\delta E$ [61]. Then, the transferred energy is subtracted from the incident electron energy, and the deflection angle is calculated [16]. The electron being excited receives energy equal to the difference between the transferred energy and its ionization potential (or the energy state in the valence band plus the band gap). Its momentum is defined from the energy and momentum conservation, calculated prior to subtraction of the ionization potential. In the initial state, target electrons are considered as particles at rest.

For elastic scattering on target atoms, the CDF formalism takes into account that at low energies, an electron interacts with the collective modes of the atomic lattice – phonons, whereas at high energies it scatters on a frozen structure of atoms, reducing to scattering on isolated atoms with increase of the electron velocity [43,62,63]. Additionally, we take into account that with increase of the electron velocity, the target atoms appear less and less screened by its own electrons. The effective charge of a target atom is set using the following approximation based on the Barkas effective charge (Eq.(10)):

$$Z_t = 1 + (Z_{a-1})\left[1 - \exp\left(-\frac{v}{v_0}Z_{a-1}^{-\frac{2}{3}}\right)\right], \quad (11)$$

where $Z_{a-1}=Z_a-1$, with $Z_a$ being the average atomic number of target atoms in a compound; $v$ is the incident electron velocity. The rationale for the shape of Eq.(11) is that the effective charge in an elastic scattering of electrons on target atoms reduces to 1 for slow electrons, thereby restoring the known phononic limit [64], while for fast electrons, the effective charge reduces to the charge of an unscreened nucleus, reproducing the high-energy limit such as the Mott cross section (screened Rutherford) [39,65]. Thus, Eq.(11) produces a smooth connection between the two limiting cases within a unified approach. An example of such a calculated electron elastic scattering cross section in $SiO_2$ is shown in Figure 1. The coefficients for calculation of the CDF are taken from Ref. [42].

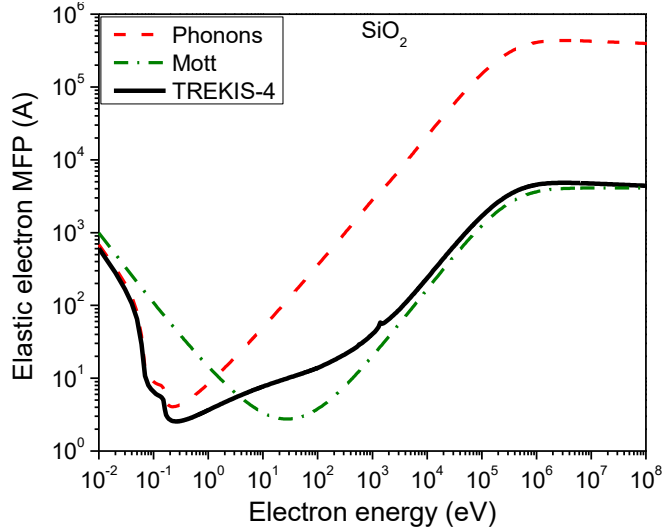

*Figure 1. An example of elastic scattering cross section of an electron in $SiO_2$ calculated with Eqs.(5),(6),(11) compared with the phononic limit and Mott's atomic scattering cross section [42].*

After the scattering, the incident electron loses amount of energy $\delta E$, calculated with Eq.(4) with the elastic cross section of scattering. This energy and coordinates of the scattering event are recorded for the bookkeeping as energy transfer to the target atoms (which will be important below, Section 2.7). The electron deflects accordingly to the energy loss, and continues its transport.

For electron Bremsstrahlung emission we use Salvat and Fernandes-Vera expression, which extends and empirically adjusts the well-known Bethe-Heitler formula [66]. All the parameters entering the formula can be found in Ref. [16]. In this scattering event, a photon is emitted, with an energy sampled from the Bethe-Heitler differential cross section according to Eq.(4). Its energy is equal to the energy loss of the incident electron. Its direction of motion is chosen along the incident electron trajectory, with no deflection produced for the electron [16].

Material surface crossing and electron emission is modelled with an Eckart-type barrier [67]. We neglect such effects as scattering of excited electron among themselves (free-free scattering), which limits the excited electron densities in the simulation to values below the densities of bound electrons. In this case, the impact ionization scattering is dominant, and free-free scattering may be neglected. We also do not account for long-range Coulomb interaction among the electrons and holes, which is also negligible for low excitation densities, close to quasi-neutral conditions, and especially for fast particles [42].

## 2.4. Holes

Holes left after impact ionizations or photoabsorption (considered below) will undergo their own kinetics, depending on the state they reside. Core shell holes will decay *via* Auger or radiative processes, emitting secondary particles. Holes in the valence band will behave like quasi-free particles, experiencing elastic and inelastic scattering events.

To model Auger and radiative decays of core holes, we use the exponential Poisson distribution of the lifetime:

$$t = -t_0 \ln(\xi_3), \tag{12}$$

Where $\xi_3 \in [0,1)$ is another random number, $t_0$ is the total decay time calculated according to the Matthiessen rule:

$$t_0 = (1/t_{Au} + 1/t_R)^{-1}, \tag{13}$$

here $t_{Au}$ is the characteristic Auger decay time of the considered shell, and $t_R$ is its radiative decay time. Both are extracted from the EPICS2017 database (EADL part), which contains the characteristic decay times for each shell of each element across the Periodic Table [52].

The shells, participating in an Auger or radiative decay, are chosen according to probabilities from EPICS2017 database [52]. After an Auger decay, a new electron is emitted with the energy equal to the difference between the binding energies of the involved shells. Its initial direction of motion is set randomly within a solid angle. The two new holes appear in the shells involved: one from which an electron fell to fill the decaying core hole, and another one an electron is emitted from.

The shell, into which a hole jumps during a radiative decay, is also chosen according to probabilities from EPICS2017 [52]. A photon is then emitted randomly within a solid angle, with the energy equal to the difference on the binding energies of the involved shells [60]. The new holes will undergo their own decays analogously, until all holes created in a cascade end up in the valence/conduction band and cannot decay anymore [42].

Energy state of a hole in the valence band is chosen according to a probability proportional to the DOS [60]. The kinetic energy of the hole is counted from the top of the valence band. The valence hole scattering is described with the same cross sections Eqs.(1-7), with its own effective mass. The effective mass of a hole is obtained from the DOS using effective one-band approximation [60].

Since holes are typically heavier than free electrons, their elastic scattering provides non-negligible energy to the atoms [68]. Inelastic scattering of valence electrons is only possible in materials, where the width of the valence band is larger than its band gap (such as semiconductors). In these events, a new electron is excited from the valence band into the conduction band, in the same manner as in the electron inelastic scattering.

Analogously to the electron transport, we neglect free-free scattering and long-range attraction or repulsion of holes.

## 2.5. Photons

Photons, either incident, or created *via* core hole radiative decay or electrons Bremsstrahlung, will experience their own scattering events. Those are: photoabsorption, elastic (Thomson or Rayleigh) scattering, inelastic (Compton) scattering, and electron-positron pair production.

Photoabsorption cross sections for each core shell of each element are extracted from EPICS2017 database (EPDL part) [52]. Photoabsorption length, $\lambda_{pa}$, for absorption by valence/conduction band electrons is calculated from the CDF using the Fano relation [45]:

$$Im\left(\frac{-1}{\varepsilon(\omega, 0)}\right) = \frac{c}{\omega \lambda_{pa}}, \quad (14)$$

In a photoabsorption event, a photon disappears, and an electron is emitted with energy $E = \hbar\omega - I_p$. Its polar angle is set randomly within the $2\pi$ interval. Its azimuth angle is set using the standard Sauter kernel [16], which makes a uniform electron emission for low-energy photons and a preferential forward direction for relativistic energies.

Elastic, also called coherent, scattering is calculated using Rayleigh expression [16]. It includes classical Thomson scattering cross section modified by the atomic form factor of the target atom. Rayleigh scattering disregards anomalous scattering factors in the current model. In such a scattering, a photon changes its direction of motion without changing its energy.

Compton scattering event is similar to the Rayleigh scattering, but a photon loses some energy in the scattering event, thereby emitting an electron from a target atom. It is also known as incoherent scattering [16]. We use the standard Klein-Nishina formula for evaluation of the Compton scattering cross section [16]. Only electron transitions to a free state (ionization) are allowed in such a process. Electron energy and emission angle are sampled from the differential cross section [16].

Electron-positron pair production for high energy photons is calculated according to the Bethe-Heitler model [16]. The screening functions and correction terms are taken from the PENELOPE model [16]. This process only occurs for photon energies $\hbar\omega > 2m_e c^2$, which does not realize in the cases we discuss in the present work.

## 2.6. Target atoms

To trace response of the target atoms to deposited energy, we use molecular dynamics (MD) method with velocity Verlet integration algorithm [69]. Periodic boundary conditions are used for atoms in the simulation box, assuming bulk target (no free surfaces). Depending on the symmetry of the problem, appropriate parts of the simulation cell may be cooled down with Berendsen thermostat [70]:

$$T(t + \delta t) = T_0 + (T(t) - T_0)\exp(-\delta t/\tau), \quad (15)$$

Where $T(t + \delta t)$ and $T(t)$ are the atomic temperature on the current and previous time step (before and after cooling), $T_0$ is the bath temperature (typically, room temperature, 300 K), $\tau$ is the characteristic cooling time, and $\delta t$ is the MD time step. Note that Eq.(15) is an exact solution of the equation for the temperature rate, in contrast to the linearized one in the original work [70], and thus it does not require the condition $\tau \gg \delta t$.

For simulation of localized ion tracks, borders of the simulation box are cooled with the Berendsen thermostat [17], whereas for laser pulse simulations, an entire supercell may be homogeneously cooled at a slower rate, or no cooling at all may be used [37].

In the current work, to simulate $Al_2O_3$, we apply Matsui potential, which is a Buckingham-type potential with a Coulomb term [71]. For evaluation of the Coulomb interaction, we apply Wolf's method of truncation in its energy-conserving form [72].

The initial conditions for atoms are set as a perfect crystalline structure with the random velocities assigned to atoms with the Maxwellian distribution (at room temperature). The atoms are allowed to equilibrate for a few hundred femtoseconds prior to arrival of radiation. The energy transfer from the excited electronic system (modeled with MC as described above) to the MD atoms is described in the next section. Atoms receive this transferred energy at each time step of the simulation *via* velocity scaling. Atomic snapshots illustrations are prepared with help of VMD software [73].

## 2.7. Energy exchange

We account for three mechanisms of energy transfer between excited electrons and atoms: elastic scattering of electrons, of holes, and nonthermal heating of atoms. The energy transfer in elastic scattering of electrons and holes, recorded in each scattering event as described in the previous sections, is averaged over all MC iterations on each time step of concurrent MC-MD simulations. It forms a spatial distribution of energy to be fed to atoms.

As was discussed in [74] (based on the theoretical results from [37,75,76] and experimental ones in [8,77,78]), nonthermal effects quickly convert potential energy built up due to electronic changes of the interatomic potential into kinetic energy of atoms. The modifications of the interatomic potential due to electronic excitation are not straightforward to include in a classical MD [18]. Although it is in principle possible to use electronic-temperature dependent interatomic potentials, in practice they are only available for a limited amount of solids [18], and are tricky to implement [79]. In the current implementation of TREKIS-4, we use a different approach.

Taking into account empirically tested idea of approximating this energy conversion as the band gap energy from each created valence hole [28], we set the following procedure. When a valence hole in the simulation loses its energy below a chosen cut off, its remaining kinetic energy and the potential energy equal to the band gap of the material ($E_{gap}$ = 8.8 eV for $Al_2O_3$) are added as the energy transferred to atoms. In this process, electrons and holes with energy dropped below the cut off disappear from the MC simulation (considered "stopped"). The cut off is set as $E_{cut}$=0.1 eV, which produces a reasonable speed of atomic heating in comparison with the *ab-initio*-based simulations of the modification of the interatomic potential and ensuing increase of the kinetic energy of atoms, as will be shown below [74].

The spatial distribution of energy lost by electrons and holes in the elastic scattering and the energy transferred when electrons and holes stop (reach the energy cut off) are averaged over all concurrently executed MC iterations at the current time step. This average energy profile is then transferred to MD atoms at the same time step of the simulation. We use a simple velocity scaling algorithm to deliver energy to atoms, which conserves the total momentum in the simulation box. More advanced methods such as Langevin thermostat may be implemented in the future [80].

The MD simulation box covers a smaller region in space than the MC grid for excited particles transport. Only the energy inside the MD box is delivered to atoms for further simulation; the energy outside of the MD box is stored for bookkeeping, to test the total energy conservation in the simulation scheme [36].

# 3. Results

Below we consider two examples. The first one is irradiation of alumina with femtosecond soft X-ray laser pulse, which will allow us to validate the developed model *via* comparison with previously published results from Ref. [81]. After validation of the model, we will consider irradiation of alumina with a swift heavy ion to gain new insights into the processes that could not have been studied before without a hybrid model resolving electronic and atomic kinetics simultaneously.

### 3.1. X-ray pulse irradiation of $Al_2O_3$

To validate the proposed combined approach, we modeled irradiation of α-$Al_2O_3$ with a femtosecond soft X-ray laser pulse: 92 eV photon energy, 10 fs full width at half maximum (FWHM) Gaussian temporal profile, and 2 eV/atom absorbed dose. To compare the results with previously published ones in Ref. [81], no thermostats were used, assuming NVE ensemble of atoms. All MC particles were locked within the simulation box equal in size to the MD simulation box with periodic boundary conditions. 1000 MC iterations are used to obtain reliable statistics. The simulation box contained 18000 atoms in TREKIS-4 simulation; in Ref. [81], the simulation box of 240 atoms was used due to computational resources consumption of the tight-binding MD of XTANT-3 code.

Applied irradiation excites initially the sample electronic system, which then deposits a part of the excess energy into the atomic one. Figure 2a shows the increase of the atomic energy due to this energy transfer. The initial increase of the total atomic energy during the first ~50 fs takes place due to interaction with the primary electron cascades, whereas the longer tail is formed by Auger-electrons, released from the *L*-shell of Al, decaying with the characteristic time of ~110 fs [52]. Figure 2a shows that the results of XTANT-3 and TREKIS-4 simulations coincide well, validating the coupling of the MC and MD modules, responsible for the energy exchange between the electronic and atomic systems.

Figure 2b shows the increase of the kinetic energy of atoms mainly via nonthermal heating in XTANT-3 (see detailed discussion in [74]). The agreement between XTANT-3 and TREKIS-4 results demonstrates that the proposed in Section 2.7 "hole termination" approximation for the conversion of the electronic excitation due to transient changes in the interatomic potential into the increase of kinetic energy of atoms captures this effect with a good accuracy. Larger oscillations in XTANT-3 results are due to finite size effects of the significantly smaller simulation box. The agreement validates the proposed simple scheme incorporating the nonthermal heating effects within a combined MC and classical MD simulations.

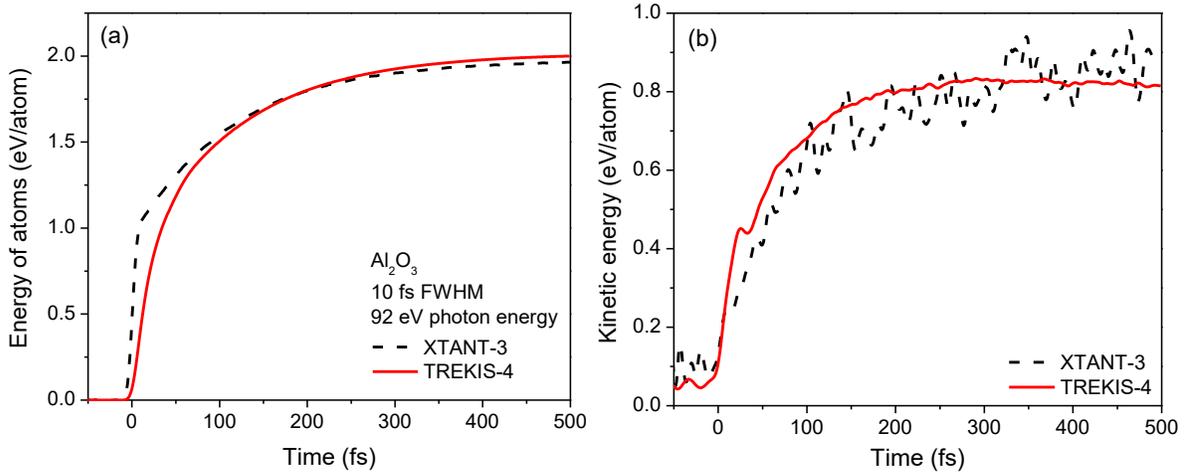

*Figure 2. TREKIS-4 and XTANT-3 simulation results of irradiation of $Al_2O_3$ with 92 eV photon energy pulse of 10 fs FWHM duration, absorbed dose of 2 eV/atom. (a) Total energy of atoms. (b) Kinetic energy of atoms.*

### 3.2. Swift heavy ion irradiation of $Al_2O_3$

To simulate a swift heavy ion impact, we use Xe ion with the incident energy of 30 MeV. Cylindrical coordinates were applied to trace energy deposition from MC particles to MD atoms. The transferred energy was traced on a radial grid with the step of 3 Å. Periodic boundary conditions for MC particles were set at the *Z* coordinate, along which the SHI was traversing once, with no boundaries along other coordinates allowing particles to travel freely. The MD simulation box containing 54000 atoms (size 12.56x14.5x2.6 nm) was used with periodic boundary conditions along all axes, and with Berendsen thermostat acting within 5 Å border layers along *X* and *Y* axes. The characteristic cooling time of 10 fs was used.

The average integral energy redistribution among various systems after an impact of Xe (30 MeV) in alumina modeled with TREKIS-4 is shown Figure 3 (for illustration, the energy is normalized to the number of atoms in the MD simulation box). This figure confirms overall energy conservation in the simulation scheme. It shows that the energy is primarily deposited by the ion to the electronic system, where holes initially possess about 50% of the energy, in agreement with the literature [68,82]. Electrons quickly lose their energy into the atomic system via both energy channels, elastic scattering and energy transfer when they "stop" – when their energy drops below the cut off. Core holes transiently accumulate a significant amount of the potential energy. They decay predominantly *via* Auger decay channel, as can be seen from a negligible amount of energy stored in photons emitted *via* radiative decays. When holes cascades are over, they end up in the valence band. Then, they release the energy into the atomic system (transferring it into the MD module) in the same manner as electrons. We see that in the case of an SHI impact, almost all the energy is transferred to atoms already within sub-100 fs timescales, also in agreement with previous studies [17,28].

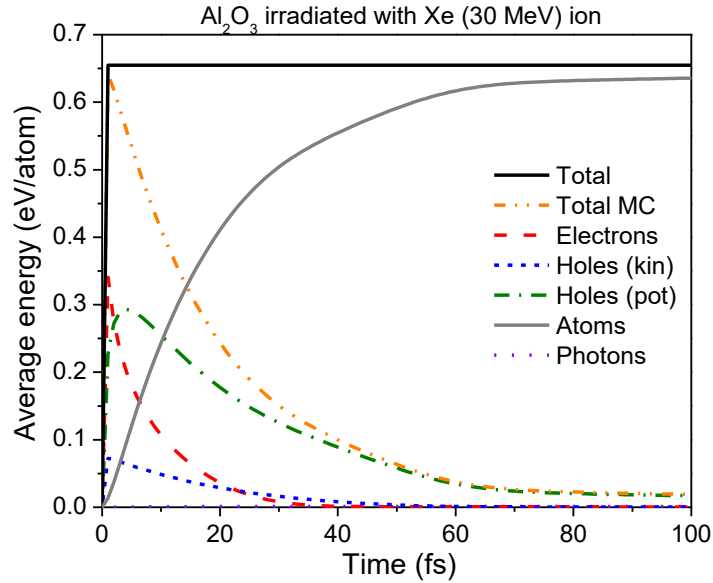

*Figure 3. Energy balance in the simulation of alumina irradiated with Xe ion (30 MeV). Total energy includes both, MC and MD simulation modules (excluding Berendsen thermostat energy drain). The curve marked as "Total MC" includes only the energy of all particles in the MC model, excluding MD atoms. Kinetic energy of valence holes and potential energy of all holes (including core holes) are shown.*

The spatial distribution of the energy transfer rate from the electronic to atomic system is shown in Figure 4. The radial distribution decays exponentially, but its shape is not constant in time. Moreover, different channels are dominant at different times. Initially, at the time of ~1 fs, the energy transfer is mostly from elastic scattering of fast electrons close to the ion trajectory. Later at ~10 fs, electrons scatter and escape outwards from the center, and form dominant contribution at larger radii. After that, electron cascades are nearly over in the proximity of the SHI trajectory – they mainly continue at large radii, not shown in Figure 4 (outside of the boundaries of the MD simulation box) [15]. Potential energy transfer (*via* band gap energy as potential energy of valence holes, see Model section 2.7) dominates later after ~10 fs. Elastic scattering of valence holes makes a small but not negligible contribution too. Since the behavior of electrons and holes is different, these sources of the energy transfer to the atomic system cannot be reliably traced with a single equation for the electronic system such as thermo-diffusion equation used in TTM-MD [31,33,83].

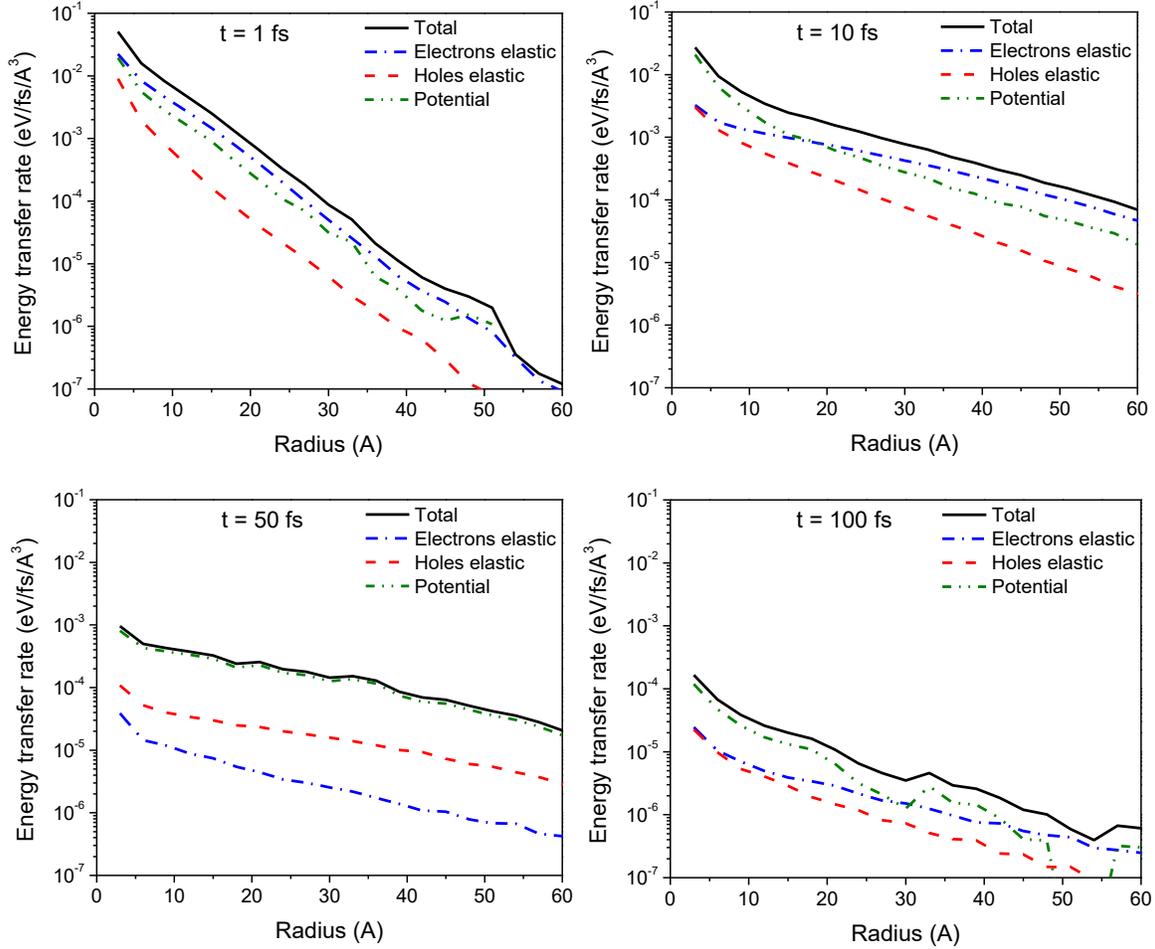

*Figure 4. Energy transfer rate from electronic (modeled with MC) to atomic (modeled with MD) system, simulated with TREKIS-4 for Al$_2$O$_3$ irradiated with Xe (30 MeV) ion at different time instants.*

Atomic snapshots shown in Figure 5 illustrate the atomic response to this energy deposition. A cylindrical track of disordered material forms around the SHI trajectory. By the time of ~50 fs, the electron cascades are nearly over (Figure 3), whereas atomic disorder is just starting (cf. Figure 5). Note that the atomic disorder takes place at the timescale of a few hundred fs, much faster than the electron-phonon coupling would have heated atoms up. This atomic heating is a result of effects of electrons as well as valence holes elastic scatterings which do not involve phonons [43,62,63], and important nonthermal atomic heating via conversion of the changes of the potential energy of atoms into their kinetic energy [74]. All the three mechanisms of the energy exchange are different from the electron-phonon coupling. They cannot be captured with the formalism of the two-temperature model, which resulted in the need to use adjustable parameters in TTM-MD models of SHI track creation [31,33,83].

The formation of a track in Figure 5 agrees well with previously reported simulations [28], as discussed in more detail in the Appendix. The results become nearly identical after ~100 fs, which proves that the model of an instantaneous energy deposition used in [28] produces reasonable results for the final track formation. However, models that use electron energy transfer as initial conditions for MD, such as Ref. [28], are missing the stage of the energy deposition into the atomic system, when the simultaneous kinetics of excited electronic system and atomic lattice response occur.

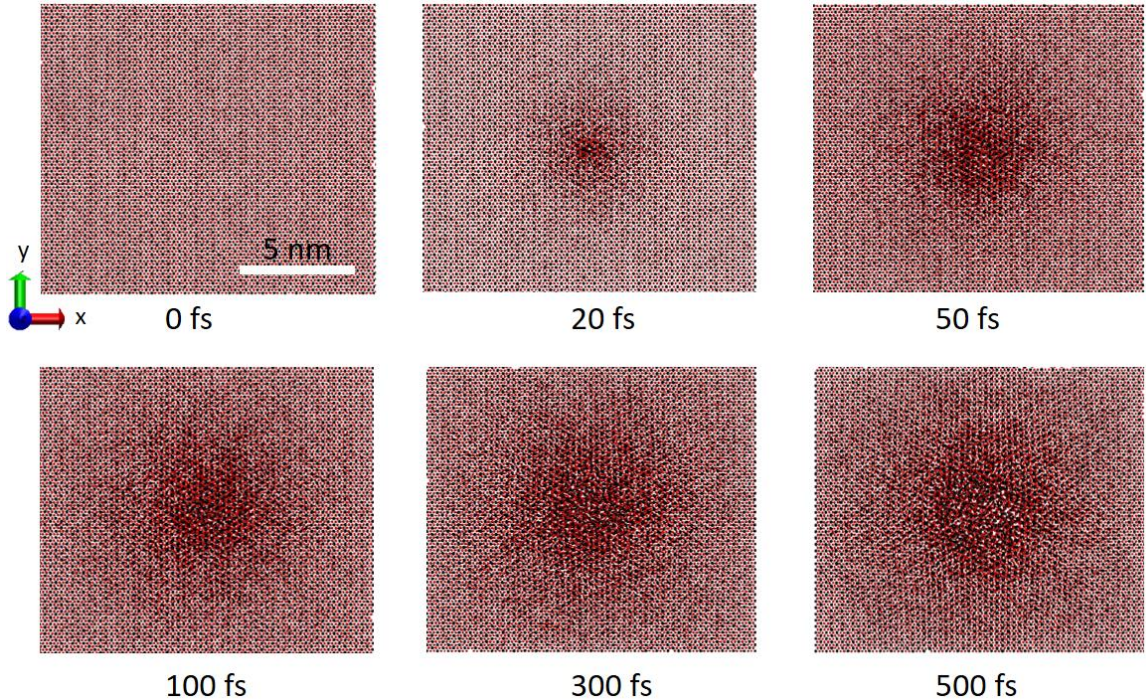

*Figure 5. Atomic snapshots of MD simulation of $Al_2O_3$ after energy deposition by Xe (30 MeV) ion, modeled with MC. Black balls are aluminum atoms, red balls are oxygen atoms.*

## 4. Discussions

Our results show that after ultrashort energy deposition into the electron system, the electronic kinetics for the most part proceeds in an unchanged atomic system. Atoms react to the energy deposition from the electrons and holes at timescales longer than those of relaxation of the electronic system. A similar notion was discussed for the case of a femtosecond X-ray irradiation before [84]. This suggests that a model, where output data of an MC simulation are used as input data for the MD simulation of atomic response, often suffices to describe the target kinetics (see Appendix for more details). It is a good approximation in the case of ultrafast energy deposition and short-lived secondary electron cascades. In this case, atomic reaction mainly takes place after the electronic kinetics is over, which means that the electronic processes mainly occur in a still unchanged material. This approximation was used in earlier works as an *a-priori* assumption [26,28,85], which could not have been confirmed without a hybrid simulation such as the one implemented in this work.

Eqs. (6) and (7) do not account for changes of the CDF with a change of a material temperature and structure. The temperature factor in the cross section (5) only account for a part of the temperature dependence [86]. For the studied cases, it is a sufficient approximation, since the cascades proceed mostly in an unirradiated part of the sample, and are over when the material properties change noticeably [15,84]. For future developments and applications of longer irradiation pulses (such as picosecond X-ray laser pulses, intensive electron or ion beams), it will be important to account for ongoing evolution of the CDF during the target response. E.g. a synergy effect of fast changes of the potential field for the atomic system with its temporal heating can considerably change electron-ion coupling parameter governing lattice excitation [74]. For that, one may consider models used in plasma physics, such as Lindhard [87], Mermin [88] or full-conserving CDF [89].

For the purposes of modeling of the final material state, MC data may even be used as initial conditions for MD, without time resolution, as was done in previous works [17,26,28,85]. Such one-way information passage significantly simplifies application of the respective models, but loses information on the initial stage of atomic dynamics and thus is unsuitable to describe time-resolved experiments [90]. The current state of the art of time-resolved ion beam experiments has a resolution of some ~500 fs [91], which is insufficient to resolve this stage. Further advances in future experiments with improved time resolution should be able to resolve it. For that, it will be crucial to describe the stage of electron-lattice energy exchange with an appropriate model, as proposed in this work. Current FEL pulses and electron diffraction experiments already have sufficient time resolution to probe this state [77,92,93]. The proposed model can be useful for such applications.

The model of conversion of the potential energy of electron-hole pairs into the kinetic energy of atoms that uses the band gap energy (section 2.7) applies to materials where the band gap fully collapses upon high electronic excitation. It is primarily covalently bonded materials, whereas ionic materials may experience only partial band gap collapse [94]. In such a case, the energy release must be modified to correspond to the band gap shrinkage, estimated with help of *ab-initio* simulations. In metallic targets, nonthermal effects manifest differently, and may be accounted for e.g. via so-called electron blast force [95] or electronic pressure [96]. For metals, it may also be necessary to include an additional model for conduction band electrons, such as thermo- or hydro-dynamical approach [85].

We have to point out a limitation of the model connected with the fact that semi-empirical MD potential was used in TREKIS-4. The tight-binding MD in XTANT-3 predicted a transient formation of a superionic state in irradiated $Al_2O_3$ [81], which was supported by the DFT-MD simulations [97]. In this state, oxygen sublattice melts, whereas aluminum one stays crystalline. TREKIS-4 simulation was unable to reproduce this phase, because the MD potential used was constructed for the ground state of electronic system, and thus cannot capture effects of the modifications of the potential due to electronic excitation [18]. It is, however, a general problem of classical MD simulations, not specific to TREKIS-4. It may, in principle, be remedied by using electronic-temperature dependent potentials [98,99], but we know of none available for $Al_2O_3$.

Another limitation of the model is the fact that the linear response approximation (Eq.(5)) relies on the first order Born approximation [43,62]. Strictly speaking, it limits its applicability to electron energies above some ~50 eV (and for heavier particles, to energies rescaled by the mass ratio correspondingly). At lower energies, further improvements are necessary. However, for high-energy particle impacts, a contribution of scattering at such low energies is rather small. For swift heavy ions, the error at energies below the Bragg peak is compensated to some degree by the usage of the effective charge, Eq.(10).

## 5. Conclusions

We conclude that the reported simulation scheme of irradiation of non-metallic materials combining Monte Carlo with molecular dynamics on-the-fly presents a viable extension to or a replacement of the standard approaches such as TTM-MD. An MC-MD combination allows to account for nonequilibrium processes in the electronic system and their effects on the atomic dynamics. It also includes ballistic transport of various types of particles (electrons, valence holes, photons). It is straightforward to construct additional models in the same spirit to cover larger spatial or longer temporal scales, if necessary.

We proposed a simple model to account for the nonthermal energy exchange between electrons and atoms in covalent materials *via* conversion of the potential energy of electron-hole pairs into kinetic energy of atoms on-the-fly. The advantage of such an approach is that it can readily be implemented into standard MC models, without complex *ab-initio* simulations of the effect. A comparison of the results obtained for a femtosecond X-ray pulse irradiation with Monte Carlo tight-binding MD simulations validated the model to a reasonable accuracy. Application of the developed MC-MD model to $Al_2O_3$ allowed to study the atomic heating stage after a swift heavy ion impact.

Our analysis suggests that even one-way information passage from MC modeling of electronic system to MD modeling of the atomic one may suffice in many cases, such as ultrafast energy deposition and short electron cascades (when electronic processes finish before significant changes in the atomic system take place). It is a typical case of swift heavy ion or femtosecond laser pulses irradiation. Even using properly recorded MC output as initial conditions for MD would work well for the purposes of tracing post-mortem material damage, but not for a description of time-resolved experiments.

## Conflict of Interest

The authors declare no conflict of interests, financial or otherwise.

# Acknowledgements

Computational resources were supplied by the project "e-Infrastruktura CZ" (e-INFRA LM2018140) provided within the program Projects of Large Research, Development and Innovations Infrastructures. This work has been carried out in part using computing resources of the Federal collective usage center Complex for Simulation and Data Processing for Mega-science Facilities at NRC "Kurchatov Institute", http://ckp.nrcki.ru/. NM gratefully acknowledges financial support from the Czech Ministry of Education, Youth and Sports (grants No. LTT17015, LM2018114, and No. EF16_013/0001552). AEV and RV acknowledge support from the Russian Science Foundation (grant number №22-22-00676, https://rscf.ru/en/project/22-22-00676/). The work of RAR was funded by the Russian Science Foundation, The Russian Federation (grant number 20-72-00048). This work benefited from networking activities carried out within the EU funded COST Action CA17126 (TUMIEE) and represents a contribution to it.

# Appendix

Figure 6 shows a comparison of the results within first 100 fs of irradiation of $Al_2O_3$ with 30 MeV Xe ion simulated with the proposed TREKIS-4 approach vs. previous results of combined MC output data on atomic heating from TREKIS-3 as an instantaneous input for an MD simulation with LAMMPS [17,28]. The difference appears only during the stage of energy exchange between the electrons and atoms: TREKIS-4 demonstrates a smaller transient track while the energy is still being transferred to atoms vs. a larger track simulated with TREKIS-3 + LAMMPS. However, by the time when electron cascades and energy transfer are finished (~100 fs), the simulated tracks become nearly identical. A slightly larger size in the TREKIS-4 simulation at 100 fs is attributed to the used elastic scattering cross section with the effective charge of target atoms, Eq.(11), whereas TREKIS-3 used $Z_t=1$ (phononic limit).

Further relaxation is identical in both cases, since it is modelled with a classical MD with the same interatomic potential. Thus, it supports our conclusion that the final size of an SHI track may be reliably calculated with a model using appropriate initial conditions. Whereas for description of time-resolved kinetics, the proposed methodology (implemented in TREKIS-4) may be used to cover the stage where electronic and atomic dynamics overlap.

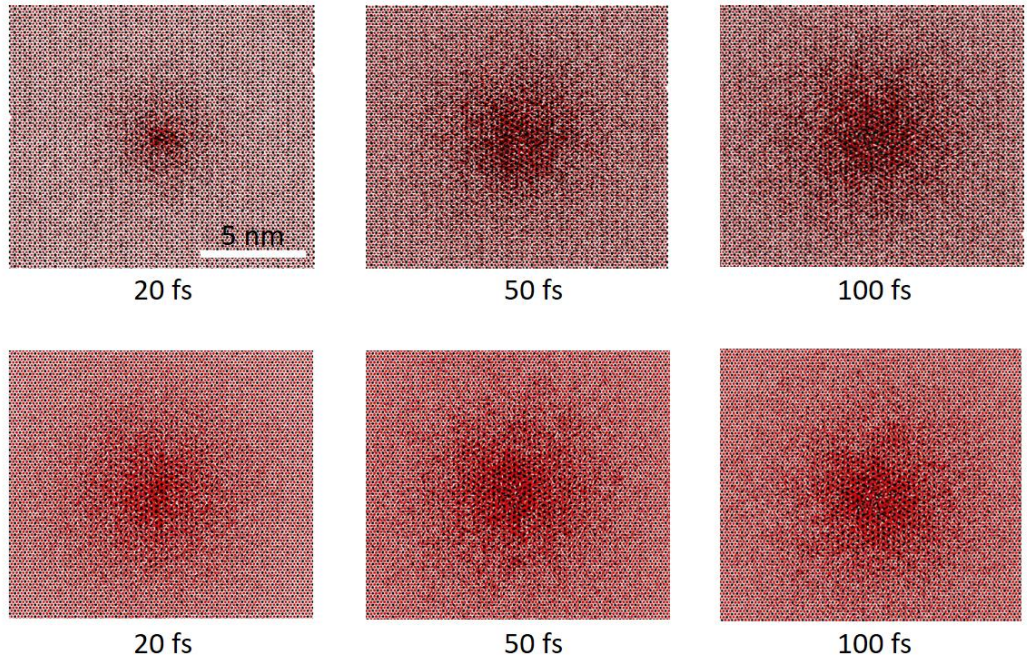

*Figure 6. Comparison between MD atomic snapshots of $Al_2O_3$ within 100 fs after irradiation with 30 MeV Xe ion, calculated with the current approach (top panels: TREKIS-4) vs. that calculated with TREKIS-3 + LAMMPS approach (bottom panels) [17,28], which uses MC output data as an instantaneous input for MD simulation.*